\documentclass[journal]{IEEEtran}
\ifCLASSINFOpdf
\else
\fi
\usepackage{multirow}
\usepackage{url}
\usepackage{float}
\usepackage{graphicx}
\usepackage[cmex10]{amsmath}
\usepackage{enumitem}
\usepackage{epstopdf}
\usepackage{amsmath}
\begin{document}

\title{Micro-location for Internet of Things equipped Smart Buildings}

\author{Faheem~Zafari,~\IEEEmembership{Student~Member,~IEEE,}
        Ioannis~Papapanagiotou,~\IEEEmembership{Senior~Member,~IEEE,} and~Konstantinos~Christidis,~\IEEEmembership{Member,~IEEE}
\thanks{Faheem Zafari is with the department of Computer and Information Technology, Purdue University, West Lafayette, IN, 47096  e-mail: \{faheem0@purdue.edu\}.}
\thanks{Ioannis Papapanagiotou is with Platform Engineering, Netflux, Los Gatos, CA, 95032 and Purdue University, West Lafayette, IN, 47096  e-mail: \{ipapapa@ncsu.edu\}.}
\thanks{Konstantinos Christidis is with IBM Emerging Technology Institute and Department of Electrical \& Computer Engineering, NC State University, Raleigh NC, e-mail:\{kchrist@ncsu.edu\}.}}


\maketitle

\begin{abstract}
Micro-location is the process of locating any entity with high accuracy (possibly in centimeters), while geofencing is the process of creating a virtual fence around a so-called Point of Interest (PoI). In this paper, we present an insight into various micro-location enabling technologies and services. We also discuss how these can accelerate the incorporation of Internet of Things (IoT) in smart buildings.
We argue that micro-location based location-aware solutions can play a significant role in facilitating the tenants of an IoT equipped smart building. Also, such advanced technologies will enable the smart building control system through minimal actions performed by the tenants.
We also highlight the existing and envisioned services to be provided by using micro-location enabling technologies.
We describe the challenges and propose some potential solutions such that micro-location enabling technologies and services are thoroughly integrated with IoT equipped smart building.
\end{abstract}


\IEEEpeerreviewmaketitle
\section{Introduction}
The developments in the field of Information and Communication Technologies (ICT) have resulted in the widespread use of reliable and affordable communication services such as the Internet.
Internet of Things (IoT) is defined as the ability of various things to be connected to each other through the Internet \cite{evangelatos2012evaluating}. The number of Internet equipped devices overtook the human population in 2011 \cite{gubbi2013internet}. As of 2013, there were 9 billion interconnected devices that are poised to reach 24 billion in 2020 \cite{gsmareport}. Groupe Speciale Mobile Association (GSMA) predicts that these devices will result in a $\char36$1.3 trillion revenue \cite{telecoms} for the mobile network operators through different services such as health, utilities, automotive and consumer electronics.
IoT is a diverse field and broadly covers Machine to Machine (M2M) communication, smart grids, smart buildings, smart cities and many more. The basic motive behind IoT is to provide advanced residential and enterprise solutions through the latest technologies in an energy efficient and reliable manner without jeopardizing the service and comfort level. It is poised to highly impact the every day life and behavior of the potential users. Due to the increasing interest in IoT and its expected impact on us, the US National Intelligence Council (NIC) has included IoT in the list of ''six Disruptive Civil Technologies'' \cite{NIC2008}. NIC forecasts IoT to penetrate the everyday entities by 2025 including furnitures, home appliances, food packages and many more. The report discusses the vast horizon of opportunities that can exist in the future. For example, integrating the popular demand with the technological advancements will drive a broad diffusion of the IoT that will contribute highly to the economic development just like Internet right now. \par
IoT plays a key role in the transformation of residential and enterprise buildings to being ``smart''.
Smart buildings aim to provide solutions that are energy efficient, environment friendly, disaster manageable and comfortable. Therefore, any solution that can potentially increase the comfort level and provides the fore-mentioned services can be incorporated into smart buildings. Indeed it is a system that allows for the buildings to have  a ``brain"' \cite{snoonian2003smart} so that they can handle the human and natural disasters properly, maintain the energy expenditure (hence reducing the greenhouse gas emission) while at the same time provide the level of comfort that the tenant asks for.  Micro-location is the process of locating an entity with a high accuracy (in centimeters). Geofencing is a related concept that creates a virtual entity around any Point of Interest (PoI). Micro-location can assist in locating a tenant within an IoT equipped smart building. The position of the user can then be utilized to provide him with effective and efficient solutions. In this paper, we aim to provide a thorough survey of various micro-location enabling technologies that can assist the IoT equipped smart buildings. We discuss various micro-location enabled services that will improve the tenant experience. We argue that due to the huge proliferation of smartphones with multiple-sensors, the tenant-building interaction can be optimized for an enhanced user experience through the utilization of micro-location enabling technologies and provision of micro-location enabled services.
We also highlight some of the challenges that micro-location is currently facing and propose effective solutions. In summary, our work presents the following concepts:
\begin{itemize}
	\item A thorough survey of various micro-location enabling technologies;
	\item Current and envisioned micro-location enabled services that can enhance the tenant's experience;
	\item Various challenges of incorporating micro-location in IoT equipped smart buildings and possible general solutions that can address the challenges;

\end{itemize}
The paper is structured as follows: Section II provides the basic description of Smart Buildings and Internet of Things. Section III presents a description  of various micro-location enabling technologies. Section IV describes the micro-location enabling techniques.
Section V presents an insight into various micro-location enabled services that are currently provided and envisioned to be provided in future.
Section VI highlights the current challenges and some of the suggested solutions, and we conclude this paper in section VII. A variation of the paper has also been publish in \cite{7120085}.

\section{IoT and Smart Buildings}
\subsection {Smart Buildings}
The Institute for Building Efficiency \cite{IBE} defines smart buildings as the buildings that can provide low cost services such as air conditioning, heating, ventilation, illumination, security, sanitation and various other services to the tenants without adversely affecting the environment.  The basic motive behind the construction of smart buildings is to provide the highest level of comfort and efficiency. For example, once a tenant enters an enterprise the temperature, humidity and the lighting are adjusted according to his personalized levels of comfort, his computer and the corresponding applications are turned on \cite{snoonian2003smart}. At the same time, the interconnection of the automation systems can assist with the disaster management and provide emergency services. For example, the fire sensors can alert the ventilation system to turn of the fans hence the smoke and the fire can be contained in a specific area. The damages in the attack on Pentagon in 2001 were reduced thanks to the advanced automation system (smart building) \cite{snoonian2003smart}.

In order to do so, there is a need for added intelligence that starts from the design phase until the building gets functional.
Smart buildings utilize IT for interconnection of various subsystems (usually independently operated).
Such interconnection results in the sharing of information that optimizes the performance of the building, allows the building to interact with the tenant, and even be connected with other adjacent smart buildings.
At the same, as the IoT is integrated into the smart building, there is a need to store, process and analyze the information obtained
from the interacting entities (tenants, other buildings, sensors etc.).
A smart building also has some level of energy independence. It has to have its own power generation through renewable energy resources, and incorporate energy efficient technologies. For example, solar photovoltaic windows can collect energy \cite{yin2012case}. At the same time, the smart building is connected with the smart grid, hence the excess energy can be provided to the grid or other buildings based on the established agreements. This may result in an extra revenue for the building. Hence, the general characteristics of a smart building are
\begin{itemize}
\item Various interconnected business systems;
\item  Equipping the tenants or people with technology;
\item Connection to various other buildings;
\item Connection to the smart grids.
\end{itemize}
Figure 1 presents an insight into characteristics and components of a smart building. Various sub-systems work together and constitute a smart building.
\begin{figure*}
	\centering
	\protect\label{fig:2}
	\includegraphics[width=1.0\textwidth]{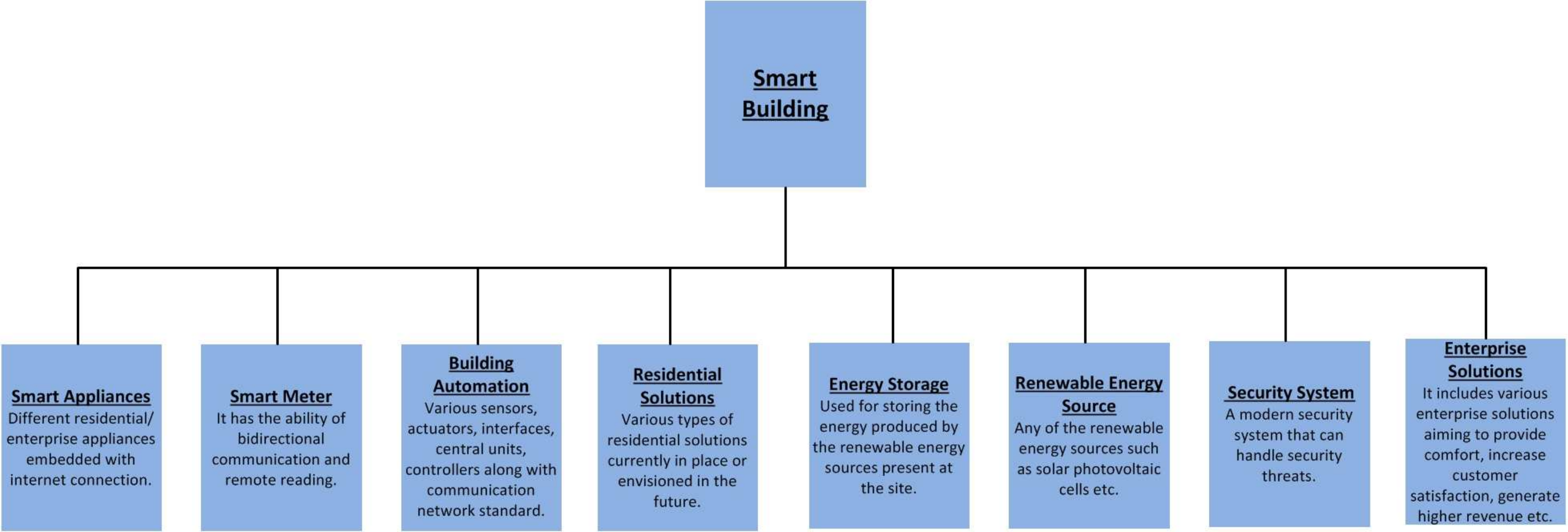}
	\caption{Various sub-systems of a smart building.}
\end{figure*}

Significant amount of research has been done to improve the architectural design of smart buildings and incorporate more technologically advanced systems into the architecture. There are two open communication standards for building automation known Building Automation and Control Networks (BACnet) and LonWorks (where Lon stands for Local Operating Network). Both these systems have a different approach towards the integration of different subsystems. BACnet being a communication only standard is mostly concerned with electrical and mechanical systems. LonWorks combines the communication part with hardware and was used in transportation and utilities industry, however it is now used in buildings as well. These two networks are not mutually exclusive and BACnet can work with some specific hardware as well. Both the systems perform their tasks differently. While BACnet defines different priority levels for performing the tasks such as the priorities of signals in the event of a fire, LonWorks might use different channels for different signals. While these two communication standards have started to establish themselves in building automation, there is still a long way to go. Such building automation systems help in reducing the operation cost \cite{snoonian2003smart}.

Wireless Sensor Networks (WSNs) are playing a key role in the smart building automation.
The WSNs are mainly designed for control and automation of smart buildings.
There are two different WSN architectural approaches, the IP based and non-IP based approaches \cite{evangelatos2012evaluating}.
The Internet Protocol Version 6 (IPv6) based Low power Wireless Personal Area Network (6LoWPAN) is the IP based system.
IPv6 is deployed over the 6LoWPAN protocol along with Constrained Application Protocol (CoAP).
The IEEE 802.15.4 Zigbee protocol serves as the link layer and the IPv6-6LoWPAN provides the services of the network or IP layer.
The application layer services are provided by the CoAP.
In the non IP based system, the physical and link level services are provided by IEEE 802.15.4, however a simpler network layer protocol can be used called Rime. It is the inbuilt IP layer of Contiki \cite{Contiki} and provides addressing as well as multi-hop networking facilities such as unicasting and broadcasting \cite{dunkels2007rime}. Rime cannot provide transport layer services, so a combined transport and application layer service is implemented via the exchange of JavaScript Object Notation (JSON) objects for the handling of application layer messages.
The IP based system performs better in terms of latency and energy consumption however the non-IP based system outperforms the IP based architecture in terms of memory footprint \cite{evangelatos2012evaluating}.

\subsection{Internet of Things}
The integration of sensing with embedded computing devices in smart buildings results in the evolution of the embedded Internet. Kevin Ashton coined the term Internet of Things (IoT) in 1999, however it was in the field of supply chain management \cite{ashton2009internet}. The term over the years has incorporated various applications that can range from health care, transportation and utilities.
The fundamental idea for IoT is the interconnection of various "Things" such as sensors, smartphones, actuators, or physical items tagged/embedded with sensors such as chemical containers with temperature sensors.. The cooperation among these devices forms the basic pillar upon which IoT stands and makes it possible for them to achieve the common goals \cite{giusto2010internet}.
The IoT is becoming extremely popular as the community looks into the possibilities from the generation of data from simple static IoT objects (e.g. a coffee machine) to the mobile devices which come with sensing, computing and communication capabilities. IoT is strongly tied to the Big Data era due to the enormous data that the "Things" can generate.
For the interconnection of these devices, different wired or wireless standards exist \cite{vermesan2013internet}. Some of the common wireless standards that are used for IoT are presented in Table \ref{tab:1}.

\begin{table*}
	\centering
	\caption{Key Wireless Technologies used in IoT}
	\label{tab:1}

	\begin{tabular}{  |p{5cm}|p{3cm}|p{4cm}|p{4.2cm}|  }
		\hline
		\textbf {Technology} & \textbf {Maximum Range} & \textbf{Maximum Frequency} & \textbf {Maximum Throughput}  \\ \hline
		\textbf {Bluetooth IEEE 802.15.1 $\textsuperscript{ \cite{ergen2004zigbee}}$} & upto 100m for Class A & 2.4 GHz & 24.0 Mbps   \\ \hline
		\textbf {ZigBee IEEE 802.15.4 $\textsuperscript{ \cite{gutierrez2001ieee}}$} & upto 100m  & 2.4 GHz & 2-250.0 Kbps   \\ \hline
		\textbf{IEEE 802.11a $\textsuperscript{ \cite{Intelwirless,ipapapa_globecom_2007}}$} & 5000m outdoor & 5 GHz & 54 Mbps \\
		 \textbf{IEEE 802.11b} & 140m outdoor & 2.4 GHz & 11 Mbps  \\
		\textbf{IEEE 802.11g} & 140m outdoor & 2.4 GHz & 54 Mbps \\
		\textbf{IEEE 802.11n} & 250m outdoor & 2.5/5 GHz & 600 Mbps \\
		\textbf{IEEE 802.11ac} & 35m indoor & 5 GHz & 1.3 Gbps with 3 antennas and 80 MHz \\
		\textbf{IEEE 802.11ad} & couple of meters & 60 GHz & 4.6 Gbps \\
		 \hline
		\textbf {WiMAX $\textsuperscript{ \cite{ipapapa_wimax_2009}}$} & depends on cell  & 2.3 GHz, 2.5 GHz and 3.5 GHz & 365 Mbps downlink/376 Mbps uplink \\ \hline
		\textbf {Long Term Evolution $\textsuperscript{ \cite{sesia2009lte}}$} & depends on cell  & 2600 MHz & 300 Mbps downlink/75 Mbps uplink  \\ \hline
		\textbf {Long Term Evolution-Advanced $\textsuperscript{ \cite{ghosh2010lte}}$} & depends on cell  & 4.44-4.99 GHz & 1 Gbps downlink/500 Mbps uplink  \\ \hline
		\textbf {Ultra Wide Band $\textsuperscript{ \cite{uwbreport}}$}  & 10-20m & 10.6 GHz  & Upto 480 Mbps \\ \hline
		\textbf{Radio Frequency Identification (RFID)$\textsuperscript{ \cite{nadlerpresence}}$} & upto 200m & upto 10 GHz & upto 1.67 Gbps $\textsuperscript{ \cite{shao2010ultra}}$  \\ \hline
	\end{tabular}
	
\end{table*}

The application and services provided by the IoT can be both residential and commercial ranging from e-health \cite{atzori2010internet}, e-marketing \cite{estimote}, intelligent car parking system \cite{faheem11survey}, intelligent transportation system \cite{weiland2000intelligent, eurasip2014},  automation, and logistic services. However, new services are nowadays deployed based on the IoT, as it is fore-seen that by the year 2025, IoT will encompass most of the appliances, food packaging, documentations, furniture and many more \cite{iotappliances,NIC2008}.
For a thorough integration of the IoT into the real world, there is a need for integrating various enabling technologies.
There are several challenges that need to be addressed including the interoperability of the devices, device smartness, security, privacy, device energy consumption, device processing capability, and network addressing \cite{atzori2010internet}.   Figure 2 presents an exemplar IoT infrastructure. Various systems and devices are interconnected through the Internet. It is evident from the figure that various devices used in residential, transportation, enterprise, healthcare environments, and literally everything else in the world can be connected through the Internet. However, incorporating all these devices with the ability to connect to the Internet will require a huge number of Internet Protocol (IP) addresses. Hence the long envisioned release of 128-bit IPv6 addresses  \cite{CiscoIPv6} is becoming a need as we move towards the IoT era.
\begin{figure}
	\centering
	\protect\label{fig:1}
	\includegraphics[width=0.5\textwidth]{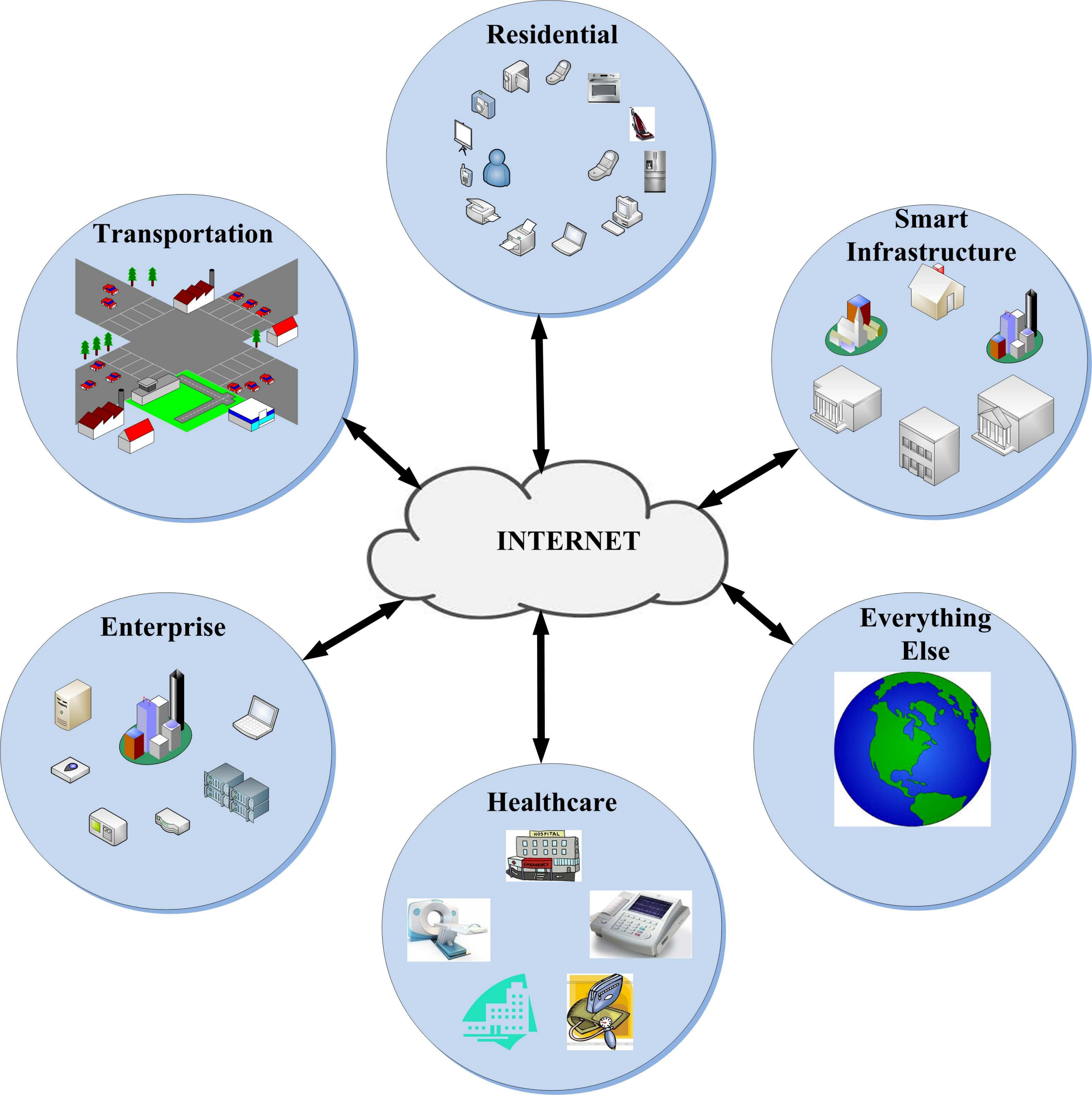}
	\caption{IoT: Connecting everything through the Internet}
\end{figure}
Due to the increase in the popularity of IoT, some research initiatives are in place now. The European Commission started pushing for IoT technologies related initiatives in 2005 \cite{yan2008internet}. National Science Foundation (NSF) in the US includes the IoT as part of their cyberphysical systems, where the goal is to design systems that merge physical and computational resources. \cite{miorandi2012internet}. However, this program also covers a wide area of applications such as smart grids, intelligent transportation, smart manufacturing and smart health care.
IoT has huge potential and a wide range applications that it can be applied to. As of now, only a very small amount of services are provided to the consumers  \cite{atzori2010internet}. The future applications are envisioned to enhance the life quality of the tenants at the office, home, gym, library, hospital etc. Particularly smart environments such as in residential or enterprise domain has a great potential. Solutions that can improve the indoor environment experience will open up lots of revenue earning sources as well as enhance the tenant comfort level. Using IoT in smart cities/smart buildings certainly can provide reliable and efficient solutions as it will allow the user to interact with the entities.

\subsection{Application Layer Technologies}
The basic feature of any application layer technology in the IoT is the fact that these devices are resource-constrained and may function in constrained IP networks. A constrained IP network may exhibit high packet loss, small packet sizes, but needs to scale to a substantial number of devices. These devices, otherwise referred as the Things, may switch several times to sleep mode, and ``wake up'' for brief periods of time. A resource-constrained device also has limited RAM, and processing capabilities. Constrained networks can occur as part of home and building automation, energy management, and the Internet of Things.

To this end, there are two IETF efforts to standardize the transactions in the IoT world.
The Constrained Restful Environment (CoRE) working group defines the framework for a limited class of applications, i.e. those that deal with the manipulation of minimal resources in constrained environments. This class includes the Things, whether they are sensors (e.g. temperature sensors, and power meters), or actuators (e.g. light switches, heating controllers, and door locks).
The second effort refers to the Concise Binary Object Representation (CBOR), a data format meant as a building block for IoT protocols. CBOR messages can be serialized using compact code to fairly small message sizes. Both of these characteristics allow for faster transmission and processing on  resource-constrained IoT nodes. The data format is based on the JavaScript Object Notation (JSON) data model, a plus given the development community's direction towards end-to-end JavaScript for Web services. CBOR allows the encoding of binary data, and is itself encoded in binary. Finally, it is extensible through a tagging mechanism that identifies data that warrants additional information.

Moreover, using IoT in smart buildings will require using a large number of sensor nodes for the provisioning of various services. The management of such large number of sensors is an issue itself. The network must have the capability of efficiently self diagnosing and self healing.
The integration of various standards results in the Building Management Systems (BMS). Open Framework Middleware (OFM) can be used for the management of WSN in smart buildings \cite{brennan2009open}.
A rule based fault analysis engine and structured knowledge when coupled with the OFM will help in root cause analysis and  the network event correlation. Such systems can be explicitly interfaced with the BMS. Such a solution is one of its kind since the management of sensors to be used for IoT in smart buildings is an uphill task and there is hardly any general purpose WSN management middleware present in the literature.
Software architectures can be used for asset management in smart buildings that can facilitate the engineers in receiving and updating the work orders and information about assets through utilization of mobile technologies and augmented reality \cite{suzuki2014smart}. A three layered architecture that contains different modules consisting of data collection, work order and asset management and event enrichment and management can help in building management system (BMS). Message Queue Telemetry Transport Protocol (MQTT) can be used for message exchange among various system components while Common Alert Protocol can be used to model the order of work and different alerts that can be then forwarded to the augmented reality application. The significance of such architecture is that the system functionality can be encapsulated and the interoperability of various subsystems is guaranteed. Also, various data standards can be integrated and the maintenance, upgrading and management of the various service requests and building assets becomes easy.

\section{Micro-location enabling technologies}
 In smart buildings due to the indoor nature, it is of primary importance to locate the user in order to enable the interaction with the rest of the interconnected things. Furthermore, the location of the user can be used to provide a wide range of novel services. Micro-location and geofencing fall in the broad category of the Location Based Services (LBS).
 LBSs have been widely used in outdoor environments for navigation services in cars, airplanes etc. For example, the widely used  Global Position System (GPS) allows a user to identify their coordinates on a map with an accuracy of approximately 10m \cite{lamarca2005place}. This fact, combined with the system's poor performance indoors (no line of sight with the satellites) and its high toll on battery life, render it unsuitable for use in a smart building setting apart from rough geofencing.
There is an  increasing interest in IoT and smart cities/buildings as depicated in figure 3 \cite{googletrend}, that can further increase by incorporation of micro-location and geofencing based services in such IoT equipped smart buildings \cite{sikeridis2017}. Realizing the significance of micro-location and geofencing based services and technologies in IoT equipped buildings, the number of micro-location and geofencing enabled products are seeing a significant increase such as:
For instance, in 2013 Qualcomm released the Gimbal \cite{gimbal} that uses the recent iBeacon protocol by Apple \cite{appleinc}. Similarly, Estimote \cite{estimote} in 2014 released a combination of Beacons and an  SDK that can be used to develop micro-location applications.
Several other big companies are moving towards this direction like Cisco \cite{ciscoiot}, as the IoT is considered by many the next "Big Thing" in the market and equipping smart buildings with IoT can provide much efficient solutions. In our study we specifically focus on the use of micro-location and geofencing techniques in IoT equipped smart buildings with emphasis on indoor area, in which IoT can be seen as a System of Interaction, which is a term recently coined by IBM \cite{ibmsystemofinteraction}. Effectively, a System of Interactive Things can enhance the performance of smart buildings and can result in efficient solutions. Therefore, from hereon we would particularly stress on micro-location and geofencing LBS as its incorporation into IoT in smart buildings is full of potential.
Geofencing is considered a Zone Based LBS \cite{bareth2010xmart}. It defines a virtual fence around a certain Point of Interest (PoI). This fence can take various geometric shapes, be it rectangular, circular, or polygonal. The goal of a geo-fence is to provide targeted services related to a predefined area. Figure 4 depicts a geofence example. There are numerous examples: for example a customer that enters a museum is notified about the proper path towards the exhibition of his interest, or an Electric Vehicle that is close to the geofence of a charging facility can be notified about a discount coupon. In micro-location, the goal is to have the location of a user or object pinpointed with the highest degree of accuracy possible \cite{faheem2017icc,faheem_arxiv_2017}. This essentially allows the system to place the user within a geofence with certainty; it also gives rise to other capabilities, such as allowing the user to position themselves within a building and track their path. Some of the enabling technologies for geofencing and micro-location are presented below.

\begin{figure}
	\centering
	\protect\label{fig:trend}
	\includegraphics[width=0.5\textwidth]{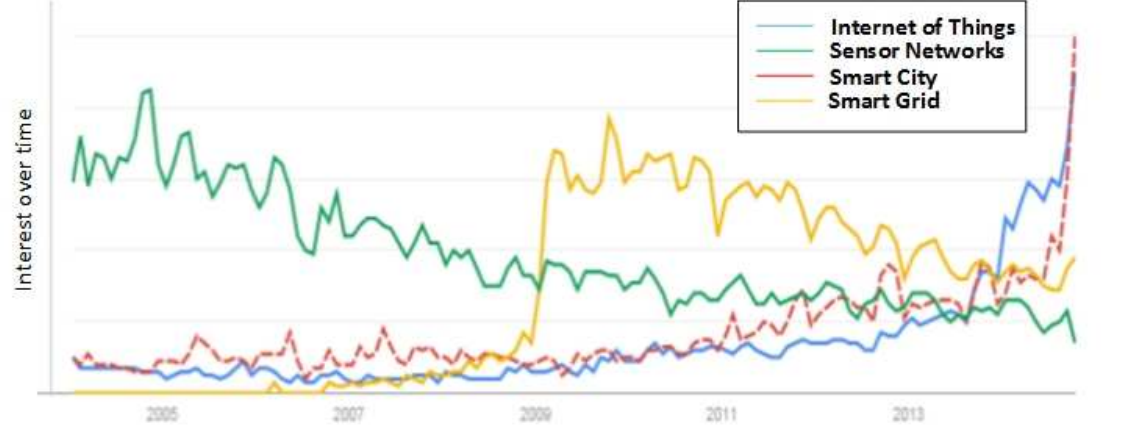}
	\caption{Google search trend for keywords IoT, smart cities, smart grids and sensor networks}
\end{figure}

\subsection{Bluetooth Low Energy based Beacons/iBeacons}
The Bluetooth Special Interest Group (SIG) proposed the Bluetooth Low Energy (BLE) or Bluetooth Smart that is also known as Version 4 of the Bluetooth Technology \cite{siekkinen2012low}.
It consists of the following layered architecture.
\begin{itemize}
\item Physical Layer (PHY): The PHY layer handles the transmission and reception of the data.
\item Link Layer (LL):  It provides the medium access, flow control and connection establishment related services.
\item Logical Link Control and Adaptation Protocol (L2CAP): It multiplexes higher layer data and provides services such as fragmenting and reassembling the large data packets.
\item Generic Attribute Protocol (GATT) and Generic Access Profile (GAP): These are the top two layers of the Bluetooth stack.
\end{itemize}
A BLE device can be either a master or a slave. A master BLE device can simultaneous connect to various slave devices however each slave is connected to a single master.In contrast to the earlier bluetooth versions,  a BLE slave advertises on either one or all three allocated advertisement channel in order to be discovered.  The Master BLE device scans the channels periodically to discover the slave devices. After the master discovers the slaves, the data is transferred through periodic connection events where both the master and device sleep and wake up to exchange the frames. The devices sleep for most of the time that enhances the energy efficiency of the devices.\par
The energy efficiency feature of BLE has made it an attractive technology for the miniscule devices known as iBeacons. The iBeacons were introduced by Apple \cite{appleinc} and the BLE based iBeacon protocol is meant to assist any BLE enabled device to detect its proximity to the iBeacon device. The iBeacon periodically transmits a beacon that can be picked up by the BLE enabled device that subsequently allows them to position themselves within the building. Due to the low energy consumption of BLE, the iBeacons can be powered through any coin cell battery which can run for years based on the configuration of the beacon parameters such as transmission power, probing frequency etc. The iBeacons can be used for both micro-location as well as geofencing. They can be used for both indoor and outdoor environments with indoor environment being the dominant one.
\par
Apple has standardized the iBeacon advertisement format \cite{passkit} and the advertising packet consists of following components.

\begin{itemize}
	
	\item Universally Unique Identifier (UUID): It is the mandatory 16 byte string that is used for identification of the ibeacons used by a specific brand or company 'A'.
	
	\item Major value: The major value is an optional 2 byte string that can be used to distinguish the iBeacons of a specific brand 'A' that are located in different localities such as a city 'B'.
	\item Minor Value: Just like major value, the minor value is also an optional 2 byte string that is used to identify the beacon of any brand 'A', in city 'B' and department 'C'.
	
\end{itemize}
The iBeacons particularly perform the task of:
\paragraph{Distance Measurement}
In order to measure the distance from a particular beacon, the BLE enabled device uses the Received Signal Strength Indicator (RSSI). The value of the RSSI is an indicator of not only the proximity of the device to the iBeacon but also shows the accuracy of the obtained estimation results.
\paragraph{Ranging}
The distance of the BLE enabled device and the iBeacon can be in any of the following four ranges \cite{appleibeaconspecification}
\begin{itemize}
	\item Immediate: The device is very close to the iBeacon.
	\item Near: The device will be in the 'near' range if it is located at a distance of about 1-3m with Line of Sight (LoS)
	\item Far: A device estimated to be far indicates that the confidence about the accuracy of the estimated proximity is low.
	\item Unknown: A device which is in the unknown range might not be close to the iBeacon or it can be due to the absence of recent initiation of ranging.
\end{itemize}
Once the BLE enabled receiver picks up the beacon from the iBeacons, it sends the specific UUID to either a server or cloud where the particular event related to the UUID is sent back to the BLE enabled receiver and is handled accordingly. It is also possible that due to obstructions, a device's range might be falsely detected.

 iBeacons is an attractive technology that can be used for micro-location purposes. Due to the expected impact of the iBeacons, it has garnered significant interest from different companies. Several companies are producing beacons and are offering beacon-based services; we list some of these in Table \ref{tab:2}.
Recently, the University of Mississippi decided to start using Gimbal's beacons for facilitating its sports fans \cite{universityofmississippi}. The basic idea is to use the beacons to enhance the game time experience of the fans and facilitate them with check in to the arena and provide them with relevant information and notifications. The beacon technology is poised to provide better consumer experience and increase the profits of the companies. The ability to provide accurate micro-location based marketing and other services can assist the consumer and be a source of great income for companies. The mobile influenced retail sales are forecasted to be 689 billion U.S. dollar in the US by the year 2016, overtaking e-commerce \cite{deloitteconsulting}. Therefore, these technologies have a great potential and their incorporation into smart buildings will facilitate both the consumer and the seller.

\begin{table*}
	\centering
	\caption{Positioning technologies and their accuracy}
	\label{tab:4}
	
	\begin{tabular}{  |p{7cm}|p{7cm} | }
		\hline
		\textbf {Technology} & \textbf {Accuracy}  \\ \hline
		\textbf {GPS} & 10m outdoor (not feasible for indoor) \cite{wang2014performance} \\ \hline
		\textbf{BLE} & Upto 10cm. \\ \hline
		\textbf{Magnetic Field Mapping } & 0.1m-2m \cite{indooratlasrange} \\ \hline
		\textbf{WLAN, Zigbee} & 3m \cite{wang2014performance}\\ \hline
		\textbf{Hybrid of RF and IR, or RF and Ultrasonic technologies} & 1m \cite{wang2014performance} \\ \hline
		\textbf{UWB based locationing} & 0.1m \cite{decawave} \\ \hline
		\textbf{RFID} & 20-30cm \cite{nadlerpresence} \\ \hline
		
	\end{tabular}
\end{table*}
\subsection{Ultra-Wideband Based Micro-location}
The Ultra-Wideband (UWB) based radio technology has a fractional bandwidth that is greater than or equal to 20\% where fractional bandwidth is the ratio of transmission bandwidth to the band center frequency  \cite{gezici2005localization,roy2004ultrawideband}. It is also defined as the wireless transmission scheme that has a an absolute bandwidth greater than 500 MHz. There are a number of advantages associated with using high bandwidth that can facilitate both communications and radars based applications. Using high bandwidth provides higher reliability because the probability of signals going around any obstacle increases due to the availability of wide range of signals having different frequencies. Also the power spectral density decreases since the signal power is spread over a large number of frequencies. Also there is a decrease in interference as well as interception probability.
For micro-location, the UWB has two different phases \cite{chong2009ranging}:
\subsubsection{Ranging}
The process of ranging involves estimation of the distance or angles between any two nodes \cite{chong2009ranging}. There are a number of techniques available for ranging such as Angle of Arrival (AOA), Received Signal Strength (RSS), Time of Arrival (TOA), or the hybrid of any these two. Since UWB has good time-domain resolution due to wide bandwidth that can provide sub-centimeter resolution ability, therefore TOA based tecniques are used with UWB for ranging.
\subsubsection{Localization}
The process of ranging results in range estimates between the fixed device (known position) and the mobile device (unknown position). The fixed device for a UWB based system is the UWB Access Point (AP) while the mobile device can be any smartphone, sensor etc.  The next step is localization i.e. estimating the location of the mobile device. There are a number of various methods available for estimating the position of the mobile device such as the Non-linear Least Square (NLS) estimator.
\par UWB technology went out of favour couple of years ago due to the complexity and various other issues, however it has recently seen a rise in interest since UWB based micro-location can provide an accuracy as high as 10cm \cite{decawave}

\subsection{Wireless Positioning Systems}
Wireless Positioning Systems are used for geofencing; they employ cellular towers for positioning outdoors, and Wi-Fi access points (AP) when considered in an indoor setting \cite{manodham2008novel}. As is the case with UWB based micro-location, some of the techniques based on WPS for positioning are angle of arrival (AOA), time difference of arrival (TDOA) and enhanced observed time difference of arrival (E-OTD). Such geometric approaches are based on the transformation of the radio frequency (RF) signal measurements into estimated distances and angles which are then used to deduce the location of the signal source using triangulation and standard geometry. The main concern associated with WPS is privacy. The cellular towers or Wi-Fi access points (AP) are vulnerable and can result in corruption of privacy. Attacks such as Man-in-the-middle attack \cite{desmedt2011man} can allow any third party to access the information by any user sent to an AP or cellular network. The user might be communicating his positioning information to the AP that the third party can have access to hence violating the user privacy. 

\subsection{Magnetic Field Mapping}
Modern day smart phones are equipped with the ability of sensing and recording the variations in the earth's magnetic field that can be used to create an indoor location map. IndoorAtlas \cite{indooratlas} is the pioneer in providing this innovative technology for finding the location of any device hence providing us with the micro-location.

\subsection{Radio Frequency Identification (RFID)}
Radio Frequency Identification (RFID) is a technology that assists in data storage and retrieval utilizing the electromagnetic transmission to some integrated circuit that is compatible with radio frequency \cite{ni2004landmarc}. The entire RFID system consists of a number of basic components such as RFID readers, tags and their intercommunication. The RFID reader is responsible for reading the data emitted by the tags. RFID systems uses specific frequency as well as protocols that govern the transmission and reception of data. The RFID tags are  categorized into
\begin{itemize}
	\item Active RFIDs: They are equipped with a battery as well as radio transceivers which enhances their range.
	\item Passive RFIDs: They operate without any battery and reflect the RF signal which is transmitted by the reader. The information is added through modulation of the reflected signal. They are meant to replace the traditional bar codes and are lighter and cheaper than their active counterparts.
\end{itemize} Despite their limited range, then can be used for positioning purposes and can be a viable option for positioning purposes environment when used in triangulation with Wi-Fi and Near Field Communication (NFC) \cite{rekha2011location}. Since most of the users rely on smart phones, the problem with RFID is that most of the devices right now do not have RFID chips or tags.  It can be used for micro-location purposes and can provide accuracy as high 20-30cm \cite{nadlerpresence}.

There are other technologies in the market that can provide positioning services such as infrared and ultrasound. Google \cite{Google}, AlterGeo \cite{Altergeo}, Skyhook Wireless \cite{skyhookwireless}, Navizon \cite{navizon}, Infosoft \cite{Infsoft}, and Combain \cite{Combain} are some of the most well-known providers of positioning services. Table \ref{tab:4} presents a summary of the discussed technologies that can provide LBS for IoT in smart buildings.

\begin{table}[ht]
	\centering
	\caption{Beacon based services}
	\label{tab:2}
	
	\begin{tabular}{  |c|p{5cm} | }
		\hline
		\textbf {Company} & \textbf {Service Description}  \\ \hline
		\textbf {Apple Inc.\cite{appleinc}} & The iBeacons are used for micro-location purposes. They are Bluetooth Low Energy devices that enable LBS. \\ \hline
		\textbf{Gimbal \cite{gimbal}} & Gimbal's beacons provides a context-aware advertising platform. The beacon's communicate through bluetooth smart and they are as per the specifications of Apple's i-beacon.\\ \hline
		\textbf{Onxy Beacon \cite{onxybeacon}} & Onxy Beacon also provides beacon based services that can help in better marketing and LBS. Onxy beacon also offers a beacon management system that is cloud based and helps in building micro location enabled applications for the beacons. \\ \hline
		\textbf{Swirl using iBeacon \cite{swirl}} & Swirl uses iBeacons and provides end-to-end mobile marketing platform within a store. It is an enterprise grade platform that helps to create, manage and optimize the LBS based mobile marketing. \\ \hline
		\textbf{Sonic Notify \cite{sonicnotify}} & Sonic Notify uses beacons to provide enterprise proximity solutions.\\ \hline
		\textbf{Estimote \cite{estimote}} & Estimote utilizes beacons for creating new and contextually rich mobile services. \\ \hline
	\end{tabular}
\end{table}

\section{Micro-location enabling techniques}
In the previous section, we covered the major technologies used for indoor localization. This section focuses on the forms of location (i.e, physical, symbolic, absolute and relative. The physical location is measured in 2D/D coordinates (e.g. degree/minutes/seconds). Symbolic location is natural expression of the location in a smart building for example, office, elevator etc. The absolute location is based on a reference grid for all located objects. The Relative location determines the proximity to a known object. Each of these techniques is independent of the technology that is used and their benefits and drawbacks are correlated with the technology in use.
\begin{figure}
	\centering
	\protect\label{fig:4}
	\includegraphics[width=0.3\textwidth]{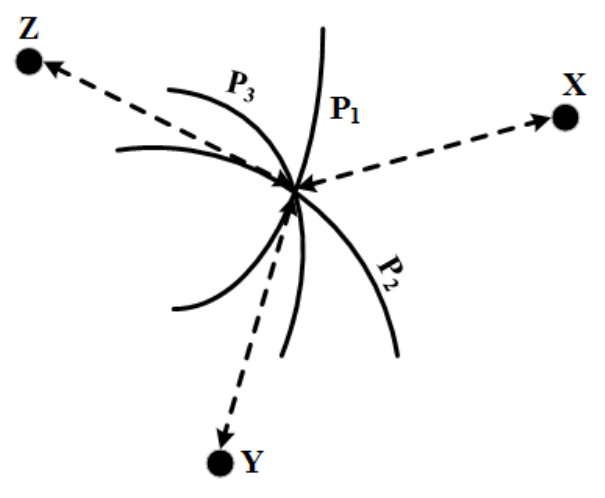}
	\caption{TOA/RTOF based localization of object.}
\end{figure}
\subsection{Triangulation}
Triangulation is the technique that involves using three dimensions to estimate the location of the target. There are two different derivations of triangulation \cite{liu2007survey}
\begin{itemize}
	\item \textit{Lateration:} the object's position is estimated through the  measurement of the distance between the object and various reference points. It is also called range measurement technique.
	\item \textit{Angulation:} relies on the computation of angles relative to a number of reference points for estimating the position of any entity.
\end{itemize}
Rather than measuring the distance directly through Received Signal Strength (RSS), the Time of Arrival (TOA), or Time Difference of Arrival (TDOA) while the distance is obtained through either the computation of emitted signal strength attenuation or multiplication of travel time and radio signal velocity.

\subsubsection{Lateration Techniques}

\paragraph{Time of Arrival}  In TOA the distance between the reference unit and target (either stationary or mobile) is proportional to the time of propagation. In order to locate any target in a 2D environment, there is need for TOA measurements with respect to the signals that are emitted by at least three reference nodes. Figure 4 shows how an object's position can be found out using three different reference points in a 2D scenario.

In TOA based systems, the propagation time (one way) is determined, and is then used to obtain the distance between the measuring unit and the signal transmitter. The problem associated with TOA is the need for precise synchronization between the systems transceivers \cite{liu2007survey}. Also there is a need for a time stamp that has to be attached to the transmitted signal so that the receiver can verify that the signal traveled directly, i.e. without being affected or reflected by any obstacle. A number of techniques such as Direct Sequence Spread Spectrum (DSSS) are used for TOA measurements.  There are number of methods to determine the position of the target. A simple method is to use the geometric method for computing the intersection points of the TOA circles. An alternative method is using the least squres algorithm \cite{fang1990simple,kanaan2004comparison} that calculates the position of any entity through the minimization of the sum of squares of the non linear cost function \cite{liu2007survey}. The assumption for such technique is that the target entity located at $(x_0,y_0) $ transmits a signal at time instant $t_0$ so the $K$ fixed stations positioned at $ (x_1,y_1),(x_2,y_2),(x_3,y_3),.....,(x_K,y_K)$ receive that particular signal emitted by the target at time instances $t_1, t_2, t_3,......, t_K$. The cost function is formulated as
\begin{equation}\label{eq:1}
C(x)= \sum\limits_{j=1}^K {\beta_j}^2 {{c_j}(x)}^2
\end{equation}
The ${\beta_j}$ depends on the signal reliability received at unit j while ${{c_j}(x)}$  can be calculated as
\begin{equation}\label{eq:2}
{c_j}(x)= v(t_j-1)-\sqrt{(x_j-x)^2+(y_j-y)^2}
\end{equation}
$v$ is the speed of light and $\textbf{x}=(x,y,t)^T$. The function can be formed for any measuring unit $j=1,2....,K$ and $c_j(x)$ can be zero using specific values of $x,y$ and $t$. The estimated location can be obtained through the minimization of $C(x)$.
\paragraph{Time Difference of Arrival}
The concept behind using TDOA is to determine the relative position of any mobile device through the examination of time difference at which a specific signal arrived at various measuring units. The difference between TOA and TDOA is that the latter relies on time difference rather than the absolute arrival time.  For every single measurement of TDOA, the transmitter must be lying on the hyperboloid with a constant difference in range between two units of measuring.  The hyperboloid equation is
\begin{align}\label{eq:3}
{R_{j,k}}= \sqrt{(x_j-x)^2+(y_j-y)^2+(z_j-z)^2} \notag \\
-\sqrt{(x_k-x)^2+(y_k-y)^2+(z_k-z)^2}
\end{align}
The $(x_j, y_j, z_j)$ and $(x_k, y_k, z_k)$ are used to represent the stationary receivers $j, k$ and $x, y, z$ represent the target's coordinates.  An easier method to solve \ref{eq:3} is to use Taylor-series expansion and formulate an iterative algorithm. Non-linear regression can also provide the exact solutions to \ref{eq:3}. Correlation techniques can also be used for computing the TDOA estimate \cite{liu2007survey}. Figure 5 shows how the location of target in 2D environment is estimated through the intersection of two or more TDOA measurements. The TDOA measurement at the points $ (X, Y, Z)$ form two different hyperbolas that can be used to locate the target W.
\begin{figure}
	\centering
	\protect\label{fig:5}
	\includegraphics[width=0.3\textwidth]{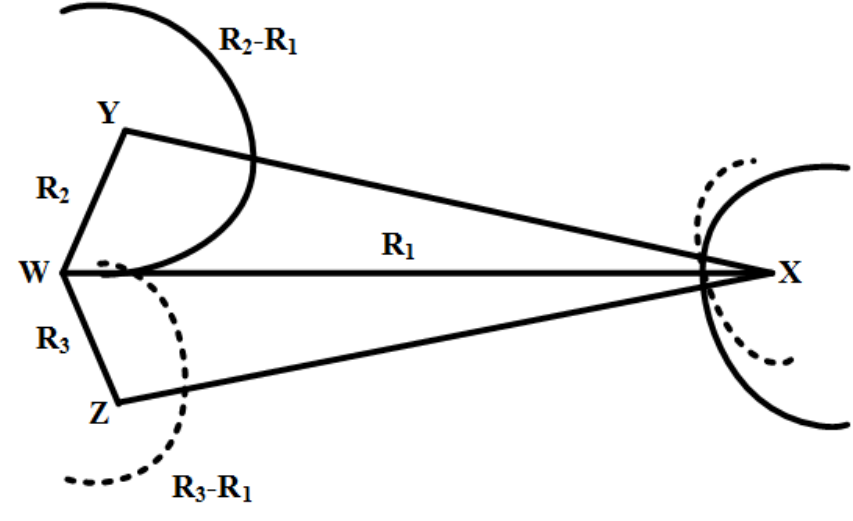}
	\caption{TDOA based localization of object.}
\end{figure}
\paragraph{Received Signal Strength (RSS) Based}
Both TDOA and TOA are affected by multipath effect since both of them rely on the time of arrival of a signal and the arrival time of a signal is itself affected by multipath effect \cite{liu2007survey}. Therefore the accuracy of the estimated position is not always great in an indoor environment. So the alternative method that is used is to estimate the mobile unit distance from a set of measuring units utilizing the emitted signal strength attenuation.  Such methods are aimed at calculating the signal path loss due to propagation.  There are a number of empirical and theoretical models that can be utilized for interpreting the difference between the transmitted and received signal strength into the estimate of range. This is shown in figure 6.

\par The path-loss models might not hold true due to the presence of extreme shadowing and multi-path fading which is the characteristic of indoor environments. The parameters of the path-loss model are site specific and the accuracy of the obtained results can be enhanced using pre-measured values of the RSS contours that are centered at the receiver.  An alternative method to improve the accuracy is through multiple measurements at a number of base stations.  Fuzzy logic algorithm can also improve the accuracy of RSS based location estimation \cite{teuber2006two}.


\paragraph{Round Trip Of Flight (RTOF)}
RTOF method relies on the measurement of round trip time of flight of any particular signal that is traveling from the transmitter to the measuring unit \cite{liu2007survey}. Figure 4 shows the concept of RTOF which is also the figure for TOA. The difference between TOA and RTOF is also that the clock synchronization requirement for RTOF is not a stringent as it is for TOA.  The mechanism of range measurement in both TOA and RTOF is the same. For both the systems, a common radar can be the measuring unit while the target replies back  to the radar signal. The complete round trip time is measured at the measuring unit. Measuring unit is again affected by the problem of knowing the exact amount of delay or processing time that is caused by the target. The delay can be ignored if it is comparatively smaller than th transmission time in a long or medium range system, but short range systems cannot ignore it.  For short range systems, modulation reflection can be a viable concept \cite{gunther2005measuring}.  It is worth mentioning that the positioning algorithms that are used for TOA can also be used for RTOF.

\begin{figure}
	\centering
	\protect\label{fig:6}
	\includegraphics[width=0.3\textwidth]{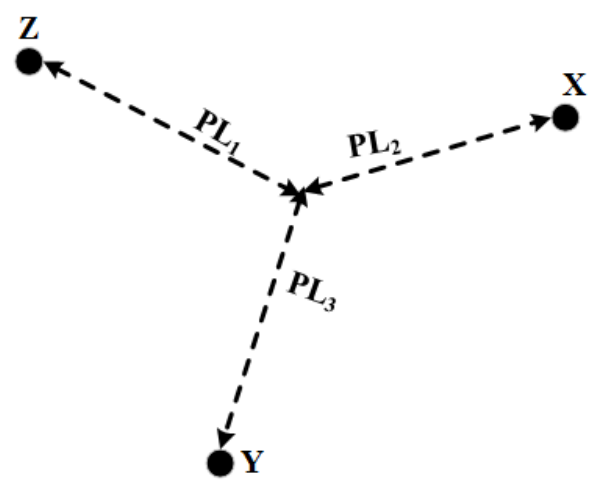}
	\caption{RSS based localization of object. $PL_1, PL_2 and PL_3$ are the path loss.}
\end{figure}
\begin{figure}[ht]
	\centering
	\protect\label{fig:7}
	\includegraphics[width=0.3\textwidth]{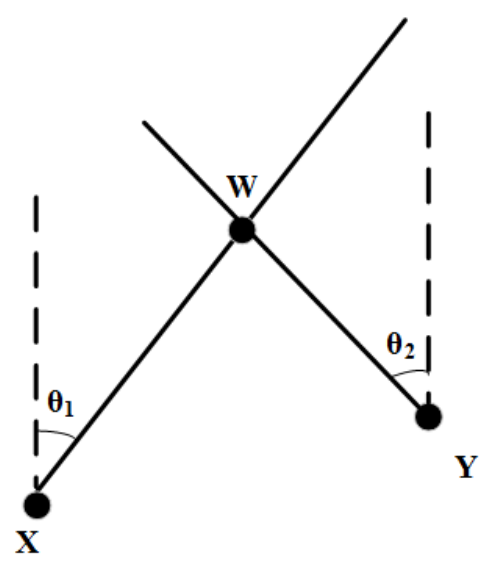}
	\caption{AOA based localization of an entity at point W.}
\end{figure}
\subsubsection{Angulation Technique}
In angulation technique such as Angle of Arrival (AoA) which is also known as direction finding the object's position is determined through the intersection of a number of pairs of angle direction lines that is formed by a circular radius of a base station to the moving target \cite{liu2007survey}. Figure 7 shows a sketch of how angulation technique can work.  For estimating the location of any entity in 2D, the method requires at least two known reference points and two angles i.e $(X,Y)$ and $ \theta_1, \theta_2$ respectively. The user's location can be estimated using AOA either through the use of array of antennas or directional antennas.

\par The advantage of AOA over other techniques is that we can use AOA to estimate the position of any entity in 3D using a minimum of three measuring units.  Furthermore, AOA does not require any time synchronization among the units of measuring. The disadvantage of using AOA is that it requires complex and large hardware and the accuracy of the estimation degrades significantly when the entity moves away from the measuring units.  In order to obtain an accurate estimate of the location of the entity, the angles must be measured accurately, however this might be impeded by the presence of multipath and shadowing in indoor environments.

\subsection{Proximity} The proximity based algorithms aim to provide information about the  symbolic relative position \cite{liu2007survey}. Such algorithms rely on using a dense grid of antennas where the position of every antenna is known. When any antenna detects a target, the target is thought to be co-located with it. However when more than one antennas detects the target, then it is assumed to be co-located with the antenna with the highest signal strength. Compared to other methods, the proximity based methods are easy to implement and its implementation is possible over a range of various physical mediums. Systems that are based on RFID and IR usually rely on proximity based methods.

\section{Micro-location enabled services}

IoT in smart buildings utilizes micro-location enabled services for various residential and enterprise solutions, in order to increase the tenants' comfort and satisfaction level. An inability to integrate such micro-location enabled services reduces the overall system efficiency and impedes innovation. Several current uses of geofences and micro-location are shown in Table \ref{tab:3}. We list some developing and envisioned scenarios below:

\begin{table*}
	\centering
	\caption{Existing geofencing and micro-location based services.}
	\label{tab:3}
	
	\begin{tabular}{  |p{5cm}|p{11cm} | }
		\hline
		\textbf {Services} & \textbf {Description}  \\ \hline
		\textbf{Fleet Management \cite{reclus2009geofencing}} & It involves managing  transportation fleets, e.g. planes, cars, vans, ships, rail cars, and trucks.  \\ \hline
		\textbf{Freight Management \cite{reclus2009geofencing}} & It involves managing the carriers belonging to a third party for ensuring the reliable, quick, and cost-effective delivery of shipments.\\ \hline
		\textbf{Mobile Tourism \cite{martin2011contextual} } & This service can assist tourists during their tours. Context-aware, LBS-based guidance and notifications can be provided. \\ \hline
		\textbf{Area-sensitive Gaming Enablement \cite{beckley2005device}} & This service enables gaming within a specific area. \\ \hline
		\textbf{Device Location and Route Adherence \cite{humphries2007method}} & It helps in locating a device and making sure the device stays within a specific location.\\ \hline
		\textbf{E-marketing} & Using geofencing and micro-location, the user can be shown location-based advertisements. It is a relatively new application, attracting most of the development interest within the smart buildings domain. \\ \hline
	\end{tabular}
\end{table*}

\subsection{Targeted E-marketing}

Geofencing and micro-location can play a great role in e-marketing. Already, there are some companies in the market that offer targeted advertising facilities that can be an added source of income.  A hardware shop owner can create a specific geofence around his shop that will send coupons, offers, and deals of the day to targeted customers who enter the geofence. Since context-awareness is accompanied with the geofence and micro-location, it can help in proper advertisement and attracting the customers to the shop. Different micro-location enabling technologies can be used for the targeted e-marketing however as of now the BLE enabled beacons are emerging as a viable option.
The beacon based context aware platforms  is the main pillar of the targeted e-marketing and they can certainly provide numerous efficient solutions.
There are numerous examples of the services that can be provided through these beacon based context aware platforms \cite{gimbal} such as:

\begin{itemize}
	\item A customer in the geofence of any retail store is notified about the special discounts.
	\item A customer notified about the availability of his prescription when he enters the geofence of a pharmacy. It is timely personalized service provided to the customer through LBS.
	\item A sports fan welcomed and given information about special discounts once he enters into a sports arena.
	\item Updating commuters pro-actively when traveling from point X to Y \cite{sanfrancisco}.
	\item A customer provided with information about a shirt that might match the shoes he just bought.
\end{itemize}

Context awareness is one of the basic requirements of targeted e-marketing since the system must know about the preference of a particular customer. In the absence of such information, the system might flooding advertisements to the customers that can lead to customer irritation. The customers can then unsubscribe to the service.
As mentioned earlier, the BLE based beacons' fundamental use as of now is targeted e-marketing and different companies \cite{estimote,gimbal,swirl} can provide such service. Although, companies have already started some of the services, they are still in infancy and there is a lot of room for improvement. The services can be further improved and can incorporate various novel concepts as well.

\subsection{Tenant Assistance}

The whole idea behind IoT and smart buildings is the facilitation of the tenant and provision of comfort and assistance to the tenants. Tenant/user assistance is a general term and can cover a wide area of services. A student who enters the library and is looking for quiet place to study can be facilitated by the smart buildings, utilizing system of interaction (Check Ref. \cite{ibmsystemofinteraction,ibmmq} for more information about system of interaction) for communicating context aware location information obtained through  micro-location and geofencing technologies. Due to the context awareness, the smart building will know that the user is looking for a quiet place. So the interconnectivity of various systems will help the building find the nearest quiet area for the user and can then provide him/her with the directions to reach the designated area. It is worth mentioning here that geofencing detected the entrance of the student into the building, micro-location found out his exact location, the context awareness helped in finding out about the student's preferred area of study while system of interaction helped in interaction of the tenant with the building by conveying obtained information and facilitating the student. All these systems working in sync with numerous other systems then facilitated the student to reach the spot. This is a classic example of how an IoT equipped smart buildings can benefit from context aware micro-location enabled services. In another scenario, geofencing and micro-location can facilitate a manager with efficient service provisions. As soon as the manager enters into the geofence of the company, his office computer and the HVAC will be turned on and the office temperature will be adjusted as per his choice using context awareness. A patient who needs emergency medical care can be facilitated by providing him with proper treatment using context awareness as well as location information for providing him immediate medical aid.  Similarly, geofencing and micro-location can be used in huge retail stores for guiding a consumer to reach a specific lane and get special discounts on his/her entity of interest. San Francisco airport is testing the beacon based micro-location system to assist the visually impaired travelers \cite{sanfrancisco}. The project is visioned to be extended to help everyone at the airport in the future, providing them with information about everything around them. In short, the range of facilities that can be provided are unlimited and it is only a matter of time until these services are provided. The use of such  micro-location services for IoT in smart buildings is certainly increasing the comfort level of the tenant and providing brand new areas of services that will result in a better standard of life. It will have a positive impact on the economy and can be a source of a great revenue.

\subsection{Energy Efficiency}

One of the driving forces behind the adoption of smart buildings is the need for energy efficient buildings. Smart buildings through the cooperation of various systems provide energy efficient solution and minimize the waste of energy.
\par

In order to obtain energy efficient smart buildings' solutions, the buildings and houses must be equipped with various capabilities such as demand side management, storage of energy on a micro level, the use of renewable energy sources on a micro level and an electricity consumption controller that relies on price signals for providing efficient solutions \cite{morvaj2011demonstrating}.
Using system of interaction and IoT in smart buildings will allow the energy consuming devices to be connected to  Internet that will  allow the user to control and monitor various appliances through a simple smartphone or any wireless terminal \cite{jeon2004remotely}. Using the micro-location enabled services, a user's location can be utilized by the energy consuming appliances that can also interact among each other and act accordingly to optimize the resources i.e. using the least possible energy to provide the optimal level of comfort to the tenant.  System of interaction facilitates the interaction of the devices and certainly will result in a paradigm shift.

\par

System of interaction Micro-location enabled services can help increase the energy efficiency of the IoT equipped smart buildings in two different ways: a) reducing the waste of energy b) optimizing the performance of the appliances and energy consuming devices. In order to provide efficient tenant assistance, the tenant must be willing to subscribe to the services. Also the tenant's position and his preferences should be used to provide the solutions properly.

\begin{figure}
\centering
\protect\label{fig:3}
\includegraphics[width=0.4\textwidth]{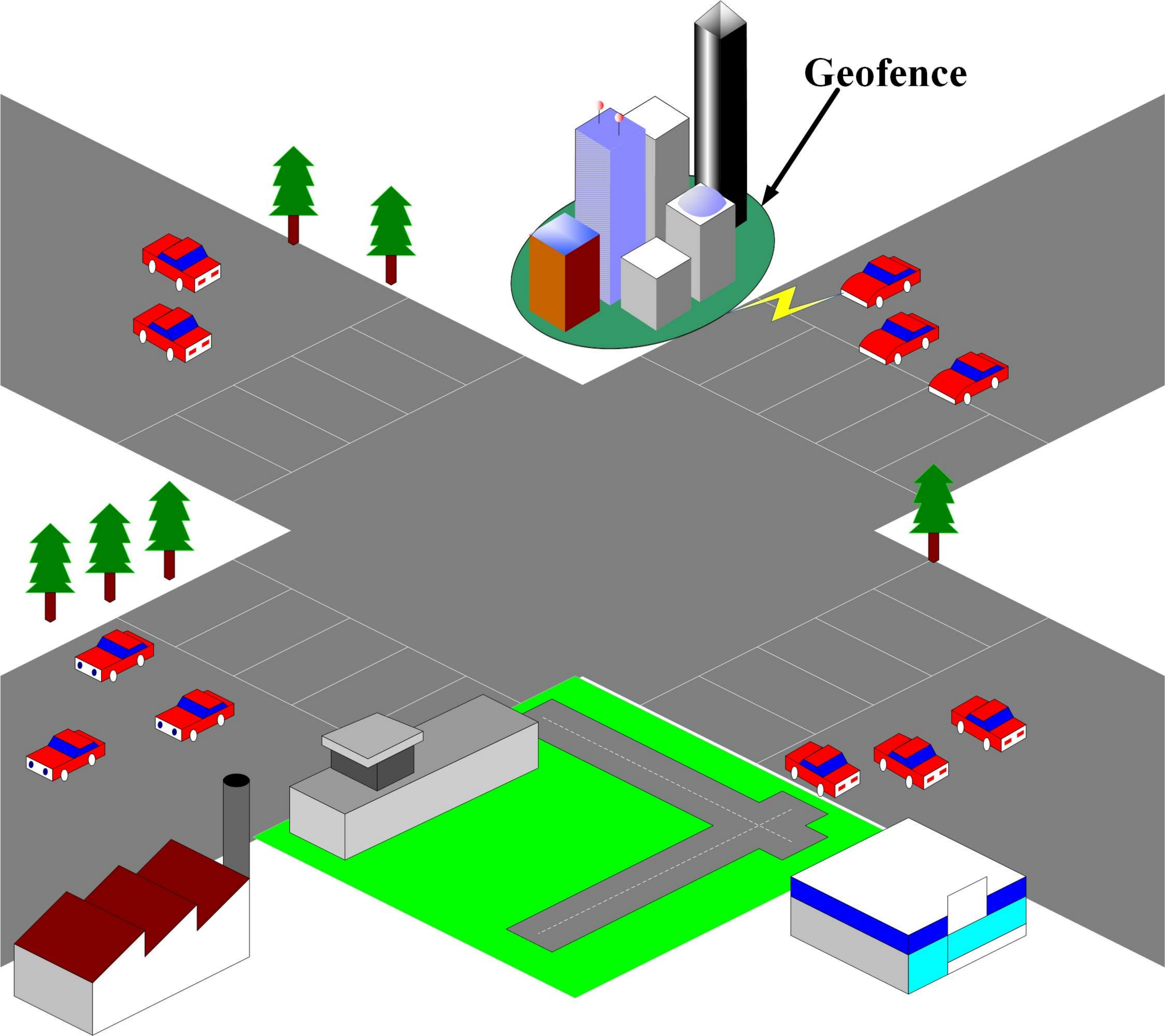}
\caption{A circular geofence around a smart building that advertises an electric charging facility to nearby electric vehicles.}
\end{figure}

\subsection{Disaster Management}

In this text, we refer to disasters inflicted by natural phenomena (floods, tornados, storms), equipment failure (e.g. fire due to a short circuit), or terrorist attacks. Traditional buildings are characterized by a low or non-existent level of preparedness for disaster management. On the other hand, smart buildings can mitigate or even completely eliminate the adverse effects of such events. For example, in the case of a fire, in additional to the fire alarm going off, the tenants will be alerted using micro-location enabled services, and the HVAC will turn off in the burning area so that the smoke cannot transfer to other parts of the building.

\par

Crowd sourcing can provide efficient disaster management solutions when it is used in smart buildings \cite{asimakopoulou2011buildings}. It can certainly provide accurate data to the disaster managers which can then be utilized for better management. Social Network Analysis can also be applied to inter link the objects for investigating and deepening the understanding of IoT paradigm \cite{zelenkauskaite2012interconnectedness}. There are a number of ways to interlink IoT that can be analyzed utilizing social network analysis. Such analysis can certainly help in disaster management and provide efficient results. The utilized smart buildings framework tend to prioritize the group safety over the safety of an individual \cite{borrion2012countering}. In order to implement such systems, it is practically and ethically required that the system should account for the uncertainty revolving around the clinical condition of every individual that can be obtained using a context aware micro-location enabled service.

\par

All of these services require high positioning accuracy, usually finer than 1m. As we've shown in Table \ref{tab:2}, UWB and BLE can provide accuracy as high as 10cm \cite{decawave}.

\section{Challenges and Proposed Solution}
The application of micro-location enabling technology and services in IoT equipped smart buildings is supposed to make things easier for the tenants, however there are certain challenges that can serve as a hurdle in the efficient utilization of micro-location enabled services that will have to be tackled for the smart buildings to serve its cause perfectly. In this section we will discuss some of those challenges and propose solutions for them.

\subsection{Interoperability}
As mentioned earlier, there are a number of micro-location enabling technologies that can be used for provision of efficient and effective solutions. These various micro-location enabling entities lack interoperability. Following sub-sections classifies the interoperability on the basis of various parameters.
\subsubsection{Technologies}
As of now, there are a number of micro-location enabling technologies available that were discussed in section III. All these technologies are different and utilizes different concepts to location with a high accuracy. IoT requires the interconnectivity and interoperability of all the various entities that constitute it. The end-user in an IoT equipped smart building is not concerned about the technologies that are used rather his position is of significant importance since that will enhance the comfort level and provide efficient services. Therefore the interoperability of different micro-location enabling technologies is of significant importance.  Currently existing micro-location enabling technologies are not inter-operable i.e. a UWB based micro-location system cannot be integrated with an ibeacon based system. The two systems based on different technologies cannot function as one unit. Similarly even within the iBeacon platform, there are different vendors that provide ibeacon based micro-location enabling services however they lack interoperability. Estimote \cite{estimote} based ibeacon cannot be detected by Gimbal's \cite{gimbal} mobile application and vice versa. This is because the vendors have their own frameworks and libraries that they use when they are developing an application. This causes lack of interoperability between the ibeacons of different vendors.

\subsubsection{Software Development Kits (SDKs)}
iBeacon vendors tend to provide their own SDKs, as a means of facilitating the development of applications. However,
SDKs of different vendors are different, and cause a vendor lock-in problem, i.e. the ibeacons of some other vendor cannot be incorporated in the beacon network as they are tied to vendor based SDK. In other words, application developers on iBeacons will have to develop different applications for their applications to function under different vendors. This also leads to issues like upgrading the iBeacons to a future generation might involve updated the whole end-to-end system.

\subsubsection{Protocols}
Another challenge is that there are not standardized protocols for micro-location enabling services. Considering again the case of iBeacons, we have observed two main protocols that are available both utilizing the BLE interface. For example,
Apple\cite{appleinc} owns the ibeacon closed-source protocol that specifies the packet and communication structure of the ibeacons and is only available to the ibeacons manufacturers. Similarly there are other vendors that provide BLE enabled beacons however they are not compatible with the Apple's standard. In such a scenario, the ibeacon will not work with the beacons that are not as per the Apple standard.  
\par Clearly, there is a need for protocols and standardization that can assist in the interoperability of technology wise different but task wise similar systems. All these devices have to work in sync to attain the required common goal i.e. to provide micro-location enabled services. Making these different technologies will also increase the overall system efficiency. 
\subsection{Privacy}

\par 
One of the main concerns with the use of micro-location enabling technologies and services is privacy. Although they are supposed to provide efficient services to the tenants of any IoT equipped building, revealing user's location is a privacy issue. As of now, the micro-location enabling technologies and services require the approval of the tenants and it is only after the tenant approves, such technologies then start operating.  Most of the tenants might find it a breach of their privacy to let their location be traced through their smart phone or any other bluetooth enabled device, therefore they might be reluctant to use such technologies and services. This is a major challenge as without tenant's consent, the micro-location enabling technologies can not reach its ultimate potential and market penetration might not be as expected.
\par In order to facilitate the growth of micro-location enabling technologies and services, the service providers must provide the tenant with a guarantee that the location information will be only used to enhance the comfort level of the tenant and will not be used for any other purpose. There is also a need for strict laws and penalties if any service provider is found violating the user privacy by using the location information for any other purpose other than the agreement with the tenant. Winning the tenant trust is an important factor in the growth of such technologies and services. 
\subsection{Energy Consumption}
Enhancing the energy efficient of the micro-location enabling technologies and services is of significant importance as such technologies and services can be energy consuming. The energy consumption for micro-location can be divided into two broad categories.
\subsubsection{Micro-location enabling devices}
The devices used for micro-location such as the ibeacons, the UWB transceivers, the magnetic field mappers and the wireless positioning system must be energy efficient. The energy consumption of such technologies can serve as a hurdle in its wide scale adoption. With the ibeacons particularly, its transmission power and transmission time period can be adjusted to save energy at the cost of performance. For energy consumption purposes, a device with lower transmission power and higher interval between the transmission of beacons is favorable.   
\subsubsection{User Devices}
The user device is one of the main component of micro-location system. No matter what specific micro-location enabling technology is used, the user device is the end device that can assist in providing the position of the user. Since the battery technology has not kept up with the pace at which the other technologies have improved, optimizing the energy consumption of the device is an important issue. The energy consumption of the devices differ based on the micro-location enabling technology used.
\paragraph{BLE enabled devices}
The BLE enabled devices due to the presence of the BLE are less energy consuming.  These devices can communicate with the ibeacons and be used for micro-location purposes. When the ibeacon used, transmits with a high time period then that will also enhance the energy efficiency of the user devices since it will have the ability to sleep for longer periods.  
\paragraph{UWB}
UWB compared to the Wi-Fi technologies provide higher bandwidth, lower power consumption but shorter range.  They are also low cost \cite{uwbreport}. However, the sstill consume a significant amoung of energy from battery limited devices, like smartphones, hence micro-location services need to take this into account.

\paragraph{RFID}
Although the passive RFIDs do not use any battery, their range is shorter. Active RFIDs due to better range needs battery power. The reliance of range on battery (due to transmission power) certainly affects the performance of the RFID. Although the smart phones as of now do not have RFID chips/tags \cite{nadlerpresence}, other user devices might be affected adversely due to the energy consumption of the RFID system. 

\par To the author's knowledge the effect of the energy consumption in micro-location services has not been studied. This is an important and open problem as these services are widely deployed and contest for minimal energy resources. Simplifying the micro-location technologies without affecting their accuracy is an interesting research problem that has to be addressed.

\subsection{Accuracy}
Since the main purpose of using micro-location enabling technologies and services in any IoT equipped smart building is to locate any user within the building to provide efficient services and solutions, the accuracy of the estimated position is of significance. Micro-location enabling technologies are supposed to have high accuracy so that the exact location of the tenant can be found out. In past, various positioning technologies such as GPS \cite{wang2014performance}, WLAN, Zigbee, Radio Frequency (RF), Infrared (IR), Ultrasounds or a hybrid of these technologies \cite{wang2014performance} have been used to find out the position of the user. These technologies can use different techniques such as RSSI, TDOA, and TOA \cite{chong2009ranging} to provide the position of the user.  These technologies are not as accurate as required for micro-location purposes (see table III). GPS is not suitable for indoor environment while the other technologies despite functioning in the indoor environments cannot attain high accuracy.  
 The accuracy range is out of the required range and there is significant room for improvement. The problem with the indoor environments is the presence of obstacles as well as various hurdles that can significantly affect the performance of the positioning technologies and techniques. Using various filtering techniques can enhance the accuracy of various technologies. Although UWB based technologies have the highest accuracy as of now \cite{decawave},  beacon based micro-location services' accuracy can be enhanced by using filters such as Kalman Filter \cite{wang2007wlan}, Extended Kalman Filter \cite{wang2007wlan} and Particle filters \cite{Zafari_globecom15,faheem2017icc} etc. There is need for further research to identify the optimal filter for micro-location and how it can further be improved to give us the best possible accuracy.   
\subsection{Security}
Although the motive behind the use of micro-location enabling technologies and services in the IoT equipped smart buildings is to facilitate that tenant with efficient and reliable solutions, there are significant security challenges that threaten both micro-location and IoT. The devices used for the purpose of micro-location are supposed to be cost-efficient and simple to minimize the power utilization, which makes them vulnerable to various attacks. All these devices can act as a point of entry for any attacker into the network therefore their security is an important issue. There is a need for authentication mechanisms.

Current authentication mechanisms rely on the binding of an identity to a pre-shared secret (e.g. a password or generated random value), a RSA key pair and its associated X.509 certificate or one-time token passwords \cite{myers1999x}. Such credentials may be prohibitive as they may be unmanned or the devices have such a small footprint, lacking in memory required to host the X.509 certificate and/or lacking in the CPU power to execute the cryptographic operations to validate the X.509 certificates (or any type of Public Key operation).

More advanced security challenges exist: as a plethora of indoor sensor, actuators and embedded systems are deployed in smart buildings, they need to adhere to a single common standard in order to facilitate zero-tough configuration and provisioning;  The scalability of micro-location enabled services in an IoT equipped smart building brings new challenges as deployments must now serve huge number of endpoints. Similarly for IoT, serving a rich multi-service edge along with all the required policies to serve the different millions endpoints forces larger and more distributed scale deployments than the classical IT. 

Such a reality teaches us that a “perfect” secure solution is unlikely to be achieved at any level. A real-time intelligent security and risk management capability provides a complementary solution to address the security gaps and threats. Hence, a flexible security framework is required. Any micro-location enabling technology or service deployment must encompass the following components:  (a) Authentication; (b) Authorization and Access Control and (c) Network Enforced Policy.

For sake of context aware services, user's information is saved on the cloud and is meant to be used for marketing purposes however it can  used for non-relevant purposes in any way that can be hazardous to the overall system.  Also due to various network attacks, precious data can be manipulated and exposed that can not only affect the seller but also the user. With the advent of beacon based LBS, a new window of marketing has opened up however its security has to be tightened and the privacy concerns of the users should be handled as well. \par

The traditional security protocols and methods cannot guarantee the security of micro-location enabling technologies and services in IoT equipped smart buildings and there is need for cutting edge research to properly secure the network.  While securing the network, it should also be made sure that the proposed solution is practically implementable on energy constrained devices. The security mechanism should be reliable as well as quick. Furthermore, those micro-location enabling technologies and services that use the cloud for data storage and other tasks face the challenges that any typical cloud based application will so the traditional security solutions applied for security of the cloud can be applied here  as well  

\subsection{Artificial Intelligence enabled devices}
Micro-location enabled services in an IoT equipped smart building are meant to enhance the tenant satisfaction and increase the overall efficiency. Due to the expected increase in future use and the overall demand of the users for intelligent and autonomous systems that will facilitate the user, there is need for enhancing the device smartness. A smart device will use user's positioning in an effective way to fulfill the tasks. In the absence of device smartness, a user might be alerted about a particular grocery to buy based on his position which might  not be the best possible option in terms of price however with the added intelligence in device, the device can find the optimum option for the user taking a number of parameters into account. This is one of the envisioned micro-location enabled services. Similarly the user's past location can be of use in predicting his future location that can be then used to provide contextual aware information with better reliability. Such services that are to be provided in an IoT equipped smart building  are challenging and require the devices to exhibit higher level of smartness  than the current level in order to accomplish their envisioned tasks in the future smart buildings. 
The use of Artificial Intelligence (AI) algorithms can certainly help in enhancing the smartness of the devices. Formulating AI algorithms that are less complex while simultaneously enhances the smartness is challenging task. Such algorithms will equip the devices to properly accomplish their tasks in the current and envisioned micro-location enabled services in IoT equipped smart buildings. Table \ref{tab:5} provides a summary of the challenges and the proposed solutions. 
\begin{table*}
	\centering
	\caption{Summary of challenges and proposed solutions.}
	\label{tab:5}
	
	\begin{tabular}{  |p{5cm}|p{5cm}|p{5cm}| }
		\hline
		\textbf {Challenge} & \textbf {Description} & \textbf{Proposed Solution}  \\ \hline 
		\textbf{Interoperability} & Different micro-location enabling technologies are not inter-operable such as UWB micro-location system cannot work in sync with an iBeacon based micro-location system & There is a need for proper standardization and protocols that can help in making different technologies inter-operable.
		 \\ \hline
		\textbf {Privacy} & The tenant's location information is sensitive and if used for any purpose other than its specific use, can result in corruption of user privacy. &Provision of guarantee to the user that his location will be only used for designated purposes as well as implementing strict laws that can punish the violators. \\ \hline
		
		\textbf{Energy Consumption} & The tenants might be reluctant to use the micro-location enabling technologies and services due to the associated energy consumption  & Develop micro-location enabling technologies and services that are highly energy efficient. Also, there is need for improving the battery technology in order to provide long lasting performance.\\ \hline
		\textbf{Accuracy} & Micro-location requires high positioning accuracy to provide better services however as of now there are some accuracy issues with micro-location enabling technologies & In order to improve accuracy, there is a need for cutting edge research. Different filtering algorithms can be used to obtain better accuracy. \\ \hline
		\textbf{Security}  & Security of the micro-location technologies and services in IoT equipped smart building can be under threat. Since some of the these technologies can use cloud, so the traditional security threats affiliated with the cloud can affect micro-location enabling technologies  & Checking the viability of distributed and centralized security mechanisms.  \\ \hline
	
		\textbf{AI enabled devices} & The smartness of the devices should be enhanced further to achieve the tasks required by micro-location enabling technologies and services in IoT equipped smart buildings  & Use of Artificial Intelligence based techniques improving the devices and enhancing their smartness.\\ \hline
	\end{tabular}
\end{table*}

\section{Conclusion}

In this paper, we focused on micro-location enabling technologies and services for an IoT equipped smart building. We described various micro-location enabling technologies that are used right now. We argued that using such micro-location enabling technologies in an IoT equipped smart buildings, we can provide the tenant with a wide range of services that will enhance the comfort level as well as increase efficiency of the overall system. We presented some of the micro-location enabled services and described some example use cases. Using the micro-location enabled services can open the door to various novel services that are only possibly due to the integration of the IoT within a smart building. Recently there have been advancements in the field of micro-location and various new technologies and techniques have been proposed. However, these advancements come with several challenges. For example,
	security, and privacy, as well as
	accuracy and energy consumption of the devices provide avenues
	for interesting research problems. To conclude, we believe that micro-location enabling technologies and services in  IoT equipped
	smart buildings have a huge potential.

\ifCLASSOPTIONcaptionsoff
  \newpage
\fi

\bibliographystyle{ieeetr}
\bibliography{references}

\end{document}